\documentclass[aps,prd,twocolumn,floatfix,nofootinbib,aps,reprint]{revtex4}
\usepackage{graphicx}
\usepackage{subfig}
\usepackage{bm}
\usepackage{ulem}
\usepackage{dcolumn}
\usepackage{hyperref}
\usepackage{cancel}
\usepackage{xcolor}
\usepackage{latexsym}
\usepackage{nicefrac}
\usepackage{latexsym}
\usepackage{amsmath, mathrsfs, bm}
\usepackage{url}
\usepackage{color}

\usepackage[percent]{overpic}

\newcommand{\er}[1]{(\ref{#1})}          
\newcommand{\dmhg}{{\rm DM}_{\mathrm{host}} }

\newcommand{\dmigm}{{\rm DM}_{\mathrm{IGM}} }
\newcommand{\dmigmm}{ \overline{{\rm DM}}_{\mathrm{IGM}} }

\newcommand{\dmcgm}{{\rm DM}_{\mathrm{CGM}} }
\newcommand{\dmcgmm}{ \overline{{\rm DM}}_{\mathrm{CGM}} }

\newcommand{\dmmw}{{\rm DM}_{\mathrm{MW}} }
\newcommand{\dmobs}{{\rm DM}_{\mathrm{obs}} }

\newcommand{\dmc}{{\rm DM}_{\mathrm{cos}} }
\newcommand{\dmcm}{ \overline{{\rm DM}}_{\mathrm{cos}} }

\newcommand{\fd}{f_{\mathrm{d}} }
\newcommand{\fdb}{\bar{f}_{\mathrm{d}} }

\newcommand{\omm}{\Omega_{\mathrm{m}} }
\newcommand{\omb}{\Omega_{\mathrm{b}} }

\newcommand{\zlim}{z_{\mathrm{lim}}}

\newcommand{\nfrb}{N_{\mathrm{FRB}}}
\newcommand{\pccm}{{\rm pc}\,{\rm cm}$^{-3}$\,}
\newcommand{\lcdm}{$\Lambda$CDM\,}
\newcommand{\tfid}{\theta_{\mathrm{fid}}}

\newcommand{\dl}{\mathrm{d}l}
\newcommand{\dz}{\mathrm{d}z}

\newcommand{\neb}{\bar{n}_\mathrm{e}}

\def\dmz{DM($z$)\,}  

\begin{document}

\title{Probing Diffuse Gas with Fast Radio Bursts}

\author{Anthony Walters$^{1,2}$, Yin-Zhe Ma$^{1,2}$, Jonathan Sievers$^{3,1}$, and Amanda Weltman$^{4}$}
\affiliation{$^{1}$School of Chemistry and Physics, University of KwaZulu-Natal, Durban 4000, South Africa}
\affiliation{$^{2}$NAOC-UKZN Computational Astrophysics Centre (NUCAC), University of KwaZulu-Natal, Durban, 4000 South Africa}
\affiliation{$^{3}$Department of Physics and McGill Space Institute, McGill University, Montreal QC H3A 2T8, Canada}
\affiliation{$^{4}$The Cosmology \& Gravity Group, Department of Mathematics and Applied Mathematics, University of Cape Town, Private Bag, Rondebosch, 7700, Cape Town, South Africa}

\begin{abstract}
The dispersion measure -- redshift relation of Fast Radio Bursts, $\mathrm{DM}(z)$, has been proposed as a potential new probe of the cosmos, complementary to existing techniques. In practice, however, the effectiveness of this approach depends on a number of factors, including (but not limited to) the intrinsic scatter in the data caused by intervening matter inhomogeneities. Here, we simulate a number of catalogues of mock FRB observations,  and use MCMC techniques to forecast constraints, and assess which parameters will likely be best constrained. In all cases we find that any potential improvement in cosmological constraints are limited by the current uncertainty on the the diffuse gas fraction, $f_{\rm d}(z)$. Instead, we find that the precision of current cosmological constraints allows one to constrain $f_{\rm d}(z)$, and possibly its redshift evolution. Combining CMB + BAO +  SNe + $H_0$ constraints with just 100 FRBs (with redshifts), we find a typical constraint on the mean diffuse gas fraction of a few percent.  A detection of this nature would alleviate the ``missing baryon problem'', and therefore highlights the value of localisation and spectroscopic followup of future FRB detections.

\end{abstract}

\maketitle

 \section{Introduction}
Fast Radio Bursts (FRBs) are a recently discovered class of brief ($\sim\mathrm{ms}$) and bright ($\sim\mathrm{Jy}$) radio transients that have been detected between $400$MHz and $1.5$\,GHz by a number of radio telescopes around the globe. Since the first detection in 2007 \cite{2007Sci...318..777L}  over 50 distinct sources have been reported \footnote{These FRB sources are available online at~\url{www.frbcat.org}.}~\cite{2016PASA...33...45P}. Their astrophysical origin remains yet unknown, and work on progenitor theories is an active field of research, with numerous different theories published to date (See review article \cite{2018arXiv181005836P} for comprehensive list of candidate theories\footnote{A progenitor theory wiki, including a summary table, can be found at {\tt www.frbtheorycat.org}.}). With a new generation of radio telescopes coming online, such as the Canadian Hydrogen Intensity Mapping Experiment  (CHIME) \cite{2014SPIE.9145E..22B}, the Hydrogen Intensity and Real-time Analysis eXperiment (HIRAX) \cite{2016SPIE.9906E..5XN}, Five-hundred metre Aperture Spherical Telescope (FAST) \cite{2011IJMPD..20..989N},  Australian Square Kilometre Array Pathfinder (ASKAP) \cite{2008ExA....22..151J}, Karoo Array Telescope (MeerKAT) \cite{2009IEEEP..97.1522J},  Murchison Widefield Array (MWA) \cite{2013PASA...30....7T}, Tianlai Project \cite{2012IJMPS..12..256C}, Deep Synoptic Array (DSA) \cite{2019arXiv190608699K}, and eventually the Square Kilometre Array (SKA) \cite{2009IEEEP..97.1482D},  expectations are high that many more samples of FRB will be detected in the near future. Initial estimates suggest the detection rate may be $\sim 39\,{\rm sky}^{-1}\,{\rm day}^{-1}$ for ASKAP~\cite{Shannon18,Liu19}, $\sim 50~\mathrm{sky}^{-1}~\mathrm{day}^{-1}$ with CHIME and HIRAX \cite{2017MNRAS.465.2286R}, and possibly as high as $\sim 10^3~\mathrm{sky}^{-1}~\mathrm{day}^{-1}$ with the SKA \cite{2017ApJ...846L..27F}. 

Owing to their excessively large Dispersion Measure (DM) at high galactic latitude, and apparent isotropy on the sky, FRBs are believed to be of extragalactic origin \cite{2015RAA....15.1629X}, possibly residing at cosmological distances \cite{2017ApJ...835...29Y,2018ApJ...852L..11S}. To date, ten FRBs have been observed to repeat \cite{2016Natur.531..202S, 2019arXiv190104525T, 2019arXiv190803507A}, one of which has been sufficiently localised on the sky for it to be associated with a host galaxy at $z\sim0.19$ \cite{2017Natur.541...58C, 2017ApJ...843L...8B, 2017ApJ...834L...7T}.  In addition, the first non-repeating burst to be associated with its host galaxy was recently reported, FRB180924 at $z=0.32$ \cite{2019Sci...365..565B}. These observations confirm that at least some FRBs propagate over cosmological distances, and this presents the possibility of using FRBs as a new probe of the cosmos. Some proposals include;  using strongly lensed FRBs to measure the properties of compact dark matter \cite{2016PhRvL.117i1301M}, or value of the Hubble constant and cosmic curvature \cite{2017arXiv170806357L}; Using a single FRB to constrain violations of the Einstein Equivalence Principal \cite{2015PhRvL.115z1101W, 2016ApJ...820L..31T},  or constrain the mass of the photon \cite{2016ApJ...822L..15W}; Using dispersion space distortions to measure matter clustering \cite{2015PhRvL.115l1301M}. All of which can be done without redshift information. Should more future FRB events be associated with a host galaxies, for which redshifts can be acquired,  this would give access to the Dispersion Measure (DM) -- redshift relation, which can potentially be used as a probe of the background cosmological parameters \cite{2014PhRvD..89j7303Z, 2014ApJ...783L..35D, 2014ApJ...788..189G, 2017A&A...606A...3Y, 2018ApJ...856...65W, 2019MNRAS.484.1637J}. 

Another cosmological puzzle that prevails through structure formation is the so-called ``missing baryon problem'', in which only $\sim70 \%$ of the baryons in the Universe can be accounted for at low redshifts \cite{2004ApJ...616..643F, 2012ApJ...759...23S}. While it is widely believed that the rest reside in the diffuse Intergalactic Medium (IGM), gaining direct observational evidence of baryon distribution is challenging. There have been a number of studies of detecting the warm-hot intergalactic medium with large-scale structure cross-correlations, such as cross-correlation between thermal Sunyaev-Zeldovich (SZ) effect and weak lensing~\cite{2015JCAP...09..046M, Hojjati15, Hojjati17}, the stacking of luminous red galaxy pairs with thermal SZ map~\cite{Tanimura19}, the detection of temperature dispersion of kinetic SZ effect within the X-ray selected galaxy clusters~\cite{Planck-dispersion2018}, and the detection of the cross-correlation between kinetic SZ effect with velocity field~\cite{Planck-unbound16,Carlos15}. Since the DM of an FRB is caused by its propagation through regions containing free electrons, they are directly sensitive to the location of baryons in the late Universe, and thus may help to shed light on where these missing baryons reside. Indeed, one such proposal considers cross-correlating FRB maps with the thermal Sunyaev-Zeldovich effect to find missing baryons \cite{2018PhRvD..98j3518M}.  Others have shown $\mathrm{DM}(z)$ data can tightly constrain the diffuse gas fraction in the IGM \cite{2019MNRAS.484.1637J}, and even possibly differentiate between that and halo gas in the CGM \cite{2019ApJ...872...88R}, though these works assume perfect knowledge of the cosmological parameters.

In this paper we revisit the constraint forecast of \cite{2018ApJ...856...65W}, while relaxing the rather strong assumption of perfect knowledge of the diffuse gas fraction ($\fd(z)$), in order to assess which model parameters are best constrained, and to determine wether FRB data may help to alleviate the missing baryon problem. To do so, we simulate a set of FRB samples out to redshift $z\simeq 3$, to generate mock DM($z$) data. We then combine the data with current constraints from CMB + BAO + SNIa + $H_0$ (hereafter CBSH), and use MCMC chain to forecast constraints on $\fd(z)$ and the cosmological parameters.  In Sect.~\ref{sec:FRB-cosmo} we review the DM measurement and of what factors it consists. In Sect.~\ref{sec:mock} we generate the mock FRB , combine with other cosmological probes, and use MCMC chain to constrain these parameters. In Sect.~\ref{sec:result} we analyze our resulting constraints. The conclusion is presented in the last section. 

\section{FRB Dispersion Measurement}
\label{sec:FRB-cosmo}
By measuring the delay in arrival time between two frequency components of an FRB one can directly infer an associated $\dmobs$, and thus the integrated column density of electrons between the observer and source. Since FRBs are believed to be of extragalactic origin, $\dmobs$ is expected to comprise of contributions from a number of distinct components of diffuse gas. These include; local contributions from the ISM and halo of the Milky Way, $\dmmw$; and non-local contributions from the intervening intergalactic medium, $\dmigm$,  circumgalactic medium (i.e.  galaxy halos), $\dmcgm$, and the FRB host galaxy, $\dmhg$.  Information about the global mean gas fraction is contained in the $\dmigm$ and $\dmcgm$ terms, while $\dmmw$ and $\dmhg$  are considered as contaminants to the signal, and so need to be removed/mitigated. We thus define the cosmic DM to be~\cite{2019MNRAS.485..648P}
\begin{align}
\dmc &\equiv \dmobs - \dmmw - \dmhg,  \notag\\
&= \dmigm + \dmcgm  \label{dme}.
\end{align}
Ultimately, the precision of the constraints derived from $\dmc$ data will be limited by the intrinsic scatter in the terms on the R.H.S. of Eq.~\er{dme}, and thus precise modeling of these systematic uncertainties is important. We neglect contribution from the interstellar medium (ISM) of intervening galaxies, as this has been show to have a minor effect on the distribution of DMs~\cite{2018MNRAS.474..318P}. 


\subsection{Mean Cosmic Dispersion}
For an extragalactic burst propagating through low density plasma, the cosmic DM in the observer's frame is given by~\cite{2014ApJ...783L..35D}
\begin{align}
\dmc= \int \frac{n_\mathrm{e}}{1+z} \dl \label{dmz},
\end{align}
where  $n_\mathrm{e}$ is the free electron density, $\dl$ is the proper length element, and the integral is calculated along the line of sight. At the level of the homogeneous Friedman-Lema\^{i}tre-Robertson-Walker (FLRW) background, the relation between proper distance and redshift is given by
\begin{align}     
\dl = \frac{1}{(1+z)} \frac{c}{H_0} \frac {  \dz }{ \sqrt{ (1+z)^3 \Omega_{\rm m} +  \Omega_{\Lambda}}  }\label{dldz},
\end{align}
where $c$ is the speed of light, $H_0$ is the value of the Hubble constant today, $\Omega_{\rm m}$ is the cosmological matter density parameter,  and $\Omega_{\Lambda}$ is the dark energy density parameter, given by the constraint equation $\Omega_{\Lambda} \equiv 1-\Omega_{\rm m}$.  The background number density of free electrons can be written as \cite{2014ApJ...783L..35D}
\begin{align}
\neb(z) = \frac{\rho_{\mathrm{c},0}  \Omega_{\mathrm{b}}  \fd(z) \chi_\mathrm{e}(z) }{ m_\mathrm{p}} (1+z^3), \label{nez}
\end{align}
where
\begin{align}
\chi_\mathrm{e}(z) = \frac{3}{4} \chi_{\mathrm{H}}(z) + \frac{1}{8} \chi_\mathrm{He}(z), \label{fe}
\end{align}
and $\rho_{\mathrm{c},0}\equiv 3H_0/ 8 \pi G$ is the critical density of the Universe today, $\Omega_\mathrm{b}$ is the cosmological baryon density parameter, $ \chi_{\mathrm{H}}(z)$ and $ \chi_{\mathrm{He}}(z)$ are the ionization mass fraction in hydrogen and hellium, respectively, and $\fd(z)$ is the sum of the mass fractions of baryons in the diffuse IGM and CGM. 

Since the free electron distribution in the late Universe is highly inhomogeneous, two sources at the same redshift will likely have significant differences in the measured  value of $\dmc$. This necessitates that one average over many sky directions in order to approach the background value~\cite{2014PhRvD..89j7303Z}. Averaging  Eq.~\er{dmz} over all angles, and using Eqs.~\er{dldz}-~\er{fe},  allows one to write the mean cosmic DM, as~\cite{2014PhRvD..89j7303Z, 2014ApJ...783L..35D, 2014ApJ...788..189G}
\begin{align}
\dmcm(z) =\frac{3 c H_0 \Omega_{\rm b}  }{8 \pi G m_{\rm p}} \int^z_0 \frac{\fd(z') \chi_\mathrm{e}(z')   (1+z') }{ \sqrt{ (1+z')^3 \Omega_{\rm m} +  \Omega_{\Lambda}}}~ {\rm d}z'  .\label{dmcm}
\end{align}
At redshifts $z\lesssim3$ hydrogen and helium are thought to be fully ionized, allowing one to set $\chi_\mathrm{H}=1 = \chi_\mathrm{He}$ there, which  gives $\chi_\mathrm{e}=7/8$. Observational constraints on $\fd(z)$ are relatively poor. A baryon census in the low-redshift Universe, from a number of different probes, yields a deficit of $\sim30\%$ when compared to the predictions of \lcdm and the CMB \cite{2004ApJ...616..643F,  2012ApJ...759...23S}. This leads to a large uncertainty in~$\fd(z)$, which could be any value between $ \sim0.5-0.9$, and evolving \cite{2012ApJ...759...23S}.  Indeed, recent results from numerical simulations suggest it decreases from near unity at $z=3$, to $\sim 0.8$ at the present time~\cite{2019MNRAS.484.1637J}. Another approach to model $\fd(z)$, based on observations, is to subtract from unity all other measured baryonic mass fractions in the Universe that do not contribute to $\fd(z)$. We thus use \cite{2019MNRAS.485..648P}
\begin{align}
\fd(z) = 1 - f_{*}(z) - f_\mathrm{ISM}(z) \label{fd},
\end{align}
where  $f_*(z)$  is the baryonic mass fraction in stars and remnant compact objects,  and $f_\mathrm{ISM}(z)$ is that in the dense ISM.

\subsection{Distribution of DMs}
Differences between sightline fluctuations of  $n_\mathrm{e}$  will cause sightline-to-sightline scatter in $\dmc$.  It is expected the primary contribution to the scatter will come from dark matter halos that are overdense in baryons, while the scatter due to fluctuations in voids, sheets and filaments of the IGM will be subdominant ~\cite{McQuinn14, 2018ApJ...852L..11S}. Since halos with mass $M_{\rm h} < 10^{10} \mathrm{M}_\odot$ are below the Jeans Mass of the IGM, and so are unlikely to be overdense in gas, only halos with mass $M_{\rm h} \geq 10^{10} \mathrm{M}_\odot$ will likely contribute to the scatter.

The distribution of $\dmc$ has been estimated  analytically \cite{McQuinn14, 2018arXiv180401548W}, as well as numerically, using N-body dark matter simulations of the cosmic web \cite{2015MNRAS.451.4277D, 2019MNRAS.484.1637J, 2019ApJ...872...88R, 2019arXiv190307630P}, or statistical methods \cite{2019MNRAS.485..648P}. A comparison of the various estimates show that the exact distribution of $\dmc$ values is sensitive to the radial gas profile of the halos, as well as spatial distribution of halos (see fig.~3 in \cite{2019arXiv190307630P} for a comparison of the most likely extragalactic DM, from various approaches).  One feature that is common among the results it that the DM distribution tends to have long-tails on the high-DM side. This is due to sightlines occasionally intersecting with a high mass halos/clusters, which induce a large departure from the background DM. 

One drawback to estimating the scatter using N-body dark matter simulations is that the resolution is often too coarse to resolve individual halos, and so would possibly underestimate their effect \cite{2019arXiv190307630P}. Recent numerical simulations which account for individual halo contributions have shown that the scatter due to halos can be very large indeed, with $\dmc$ values between $\sim800-2000$~\pccm at $z=1$, and long Poisson tails to the high-DM side (see fig.~17 in \cite{2019MNRAS.485..648P}).  Such large scatter may challenge the effectiveness of using $\mathrm{DM}(z)$ data to constrain $\fd(z)$ and the cosmological parameters. We investigate this in the following section, where we calculate the scatter due to intervening galactic halos (i.e. the distribution of $\dmcgm$ values) according to \cite{2019MNRAS.485..648P}.

\subsection{Dispersion of the Host Galaxy}
The host galaxy contribution, $\dmhg$, should depend on  a number of parameters, including host galaxy type, redshift, inclination, location of the FRB therein, as well as the FRB formation mechanism and its local environment \cite{2018arXiv180401548W}. Since there is currently no accepted FRB progenitor theory, and only three sources have been associated with host galaxies, the distribution of $\dmhg$ remains largely speculative. What is known from FRB121102, which has been associated with a star-forming dwarf galaxy at $z=0.19$, is that $\dmhg$ to can be a significant fraction of $\dmc$. After accounting for the Milky Way and IGM contributions (and uncertainties), $\dmhg$ was estimated to be between $55 - 225$ \pccm \cite{2017ApJ...834L...7T}. On the other hand, the recent association of FRB180924 and FRB190523 with low star-formation rate massive galaxies at $z=0.32$ and $z=0.66$, respectively, has shown that some bursts have a clean host environment \cite{2019Sci...365..565B, 2019arXiv190701542R}.  In case of FRB180924, the expected contributions from the Milky Way and IGM exceed $\dmobs$ by $46$~\pccm. 

We anticipate that at low redshifts the scatter in $\dmhg$ may be large, and that uncertainty may challenge the effectiveness of using DM($z$) data as a cosmic probe. If such data is to assist in furthering the pursuits of precision cosmology it will be important to mitigate against this systematic.   The recent detection of eight repeating bursts by CHIME \cite{2019arXiv190803507A} will likely allow for association with their host galaxies, and thus shed some light on the distribution of $\dmhg$  in the near future.

\subsection{An Ideal Sample}
A number of future telescope arrays will be equipped with long-baseline outrigger antennae, providing high angular resolution, and the ability to precisely localise transient events on the sky. This will not only allow for the association of FRBs with their host galaxies, but also a measure of their location therein. For example, the current design plans of HIRAX will make it capable of localising transients to within $\sim 0.03$ arcseconds \footnote{See the talk given at  the Texas Symposium of Relativistic Astrophysics 2017: \url{https://fskbhe1.puk.ac.za/people/mboett/Texas2017/Sievers.pdf}}. And indeed ASKAP has already begun to make progress localising bursts \cite{2018ApJ...867L..10M, 2018ApJ...860...73E}, and recently reported the first interferometric sub-arcsecond localisation of a non-repeating burst, FRB180924, to $4~\mathrm{kpc}$ from the centre of its host galaxy \cite{2019Sci...365..565B}. 

Such precise positional information may offer a route to mitigating the uncertainty associated with the host galaxy contribution. For example, one could build a catalogue of FRBs located on the outskirts of their host galaxies, thus minimizing  the host galaxy  contribution. Alternatively, one may also be able to mitigate the host galaxy systematics if given some prescription for calculating, and subtracting off, $\dmhg$, based on its location inside the host galaxy. Thus, in the following section we simulate FRB catalogues without Milky Way or host galaxy contributions, but include additional scatter in DM to account for imperfect subtraction of the host galaxy contribution.

Assuming FRBs can be sufficiently localised on the sky to be associated with a host galaxy, the main challenge in building a large sample of DM($z$) data will likely be attaining the redshift information. Attaining the redshifts for a catalogue with $\nfrb=10^3$ should be feasible with mid- to large-sized optical telescope and a dedicated observing programme stretched over a few years \cite{2018ApJ...856...65W}.

\section{Mock Data}
\label{sec:mock}

\begin{figure*}[thb!]
\centering
\includegraphics[width=.49\textwidth]{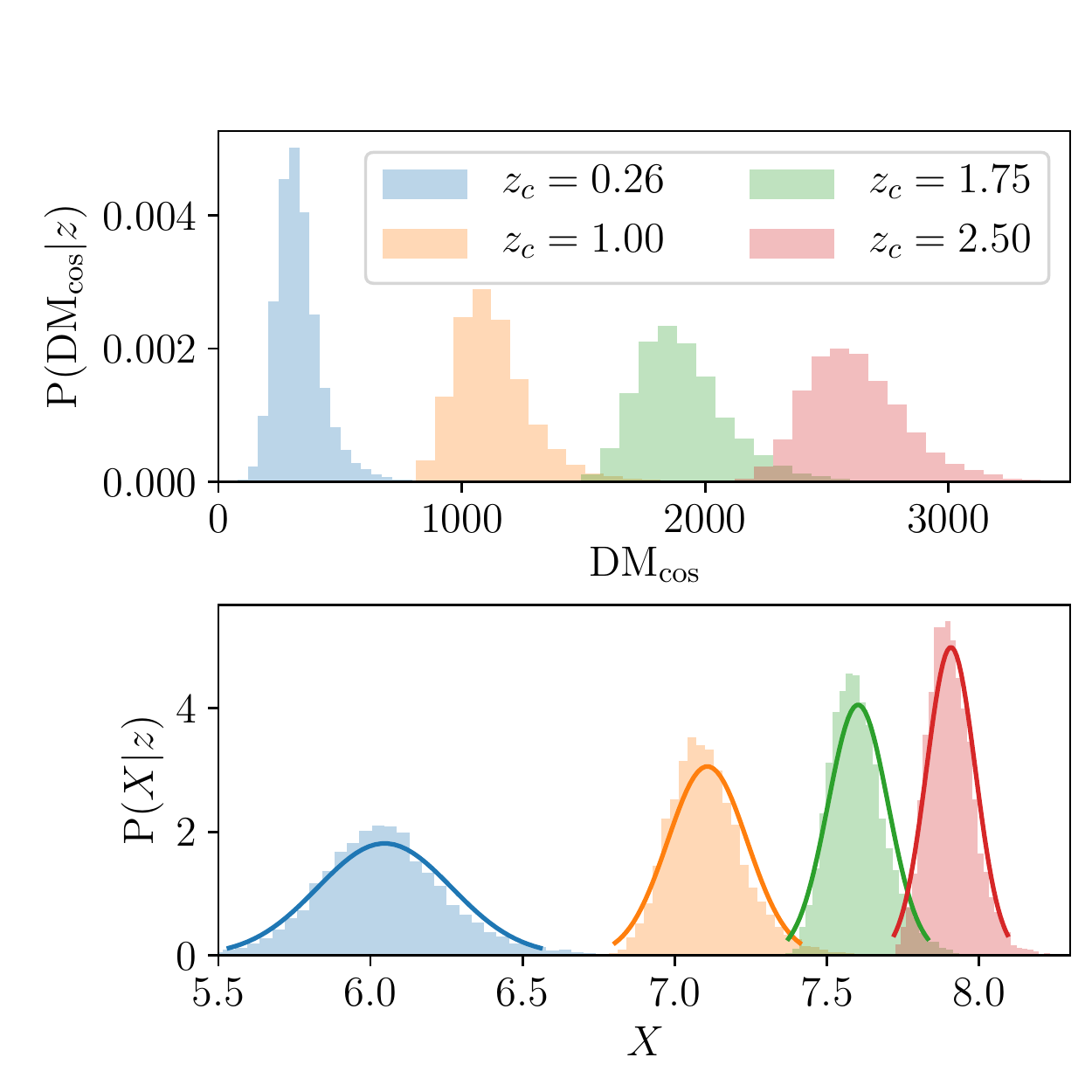}
\includegraphics[width=.48\textwidth]{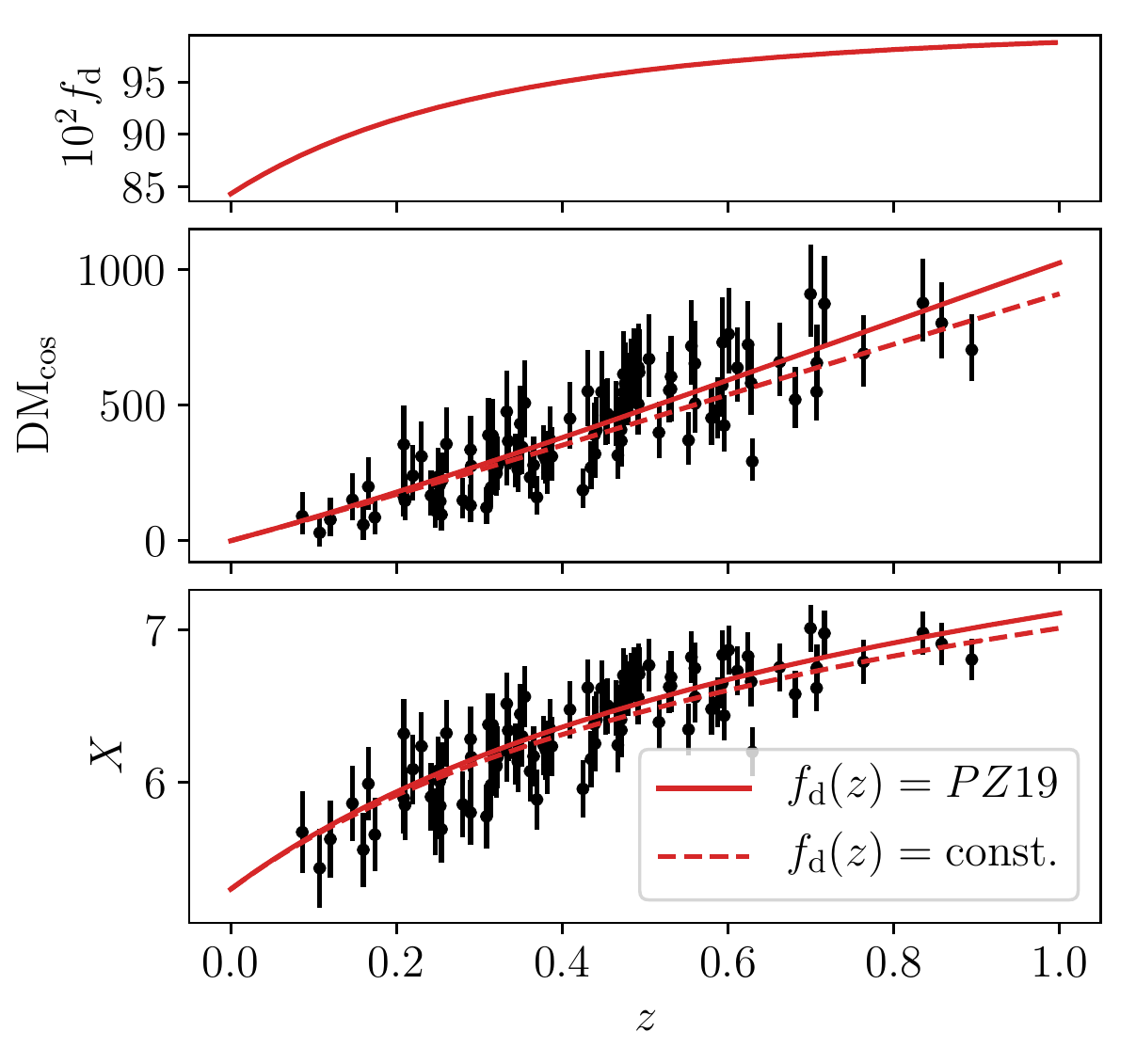}
\caption{Modeling the DM. The left panel shows the probability distributions of $\dmc$ and $X$ data  ($X\equiv\ln({\rm DM}_{\rm cos}+C)$), with coloured blocks indicating the histogram density of the simulated data in bins of $\Delta z = 0.03$, derived from $10^4$ sightlines out to $\zlim=3$. The solid coloured lines in the bottom left indicate the best-fit Gaussian to the $X$ data in that redshift bin. The right panel shows the fiducial model for $\fd(z)$ (top), and its impact on on the data (below). A single realisation of  $\dmc$ and corresponding $X$ data are shown (black points), with $\nfrb=10^2$, along with their mean values (red lines). For comparison, the mean values with constant $\fd$ are also shown (dashed red lines).  }
\label{P_DMz}
\end{figure*}

In order to generate mock FRB data we first estimate the distribution function $P(\dmc|z)$ using  Eq.~\er{dme}, while assuming a spatially-flat $\Lambda$CDM as the fiducial background cosmology, with the best-fitting CBSH parameter values provided by the {\it Planck} 2016 data release\footnote{The {\it Planck}  chains can be found at \url{http://pla.esac.esa.int/pla/\#cosmology}}, i.e. $\Omega_{\rm m}= 0.309$, $\Omega_{\Lambda}= 0.691$; $\sigma_{8}= 0.809$, and $h = 0.68$, where the Hubble constant is $H_{0} = 100 h\,{\rm km}\,{\rm s}^{-1}\,{\rm Mpc}^{-1}$~\cite{2016A&A...594A..13P}. We also assume the minimum halo mass which is overdense in baryons to be ${\rm M}_\mathrm{low}=10^{10} M_\odot$, and that the baryon profile of the halos extend out to two virial radii, $r_\mathrm{max}=2r_\mathrm{vir}$. 

To compute the contribution from the CGM (i.e. galactic halos), we simulate for $10^4$ sight-lines, out to a redshift of $z=3$, using publicly available FRB code\footnote{DM packages  can be found at \url{https://github.com/FRBs/FRB}} \cite{2019MNRAS.485..648P},  along with the {\it Aemulus Project} halo mass function emulator  \footnote{{\tt HMF\_emulator} can be found at \url{ https://github.com/AemulusProject/hmf\_emulator}}  \cite{2019ApJ...872...53M}. To each sightline we add the mean contribution of the IGM, given by 
\begin{align}
\dmigmm = \dmcm - \dmcgmm,
\end{align}
with $\dmcm$ given by Eq.~\er{dmcm}, and  $\fd(z)$ calculated according to Eq.~\er{fd}.  Expressions for $f_{*}(z)$ and $f_\mathrm{ISM}(z)$ are computed by the FRB package by fitting an interpolated spline curve to observational  stellar mass  \cite{madau} and remnant object data \cite{2004ApJ...616..643F}.  A  for a plot of the fiducial model for $\fd(z)$ see the top right panel of Fig.~\ref{P_DMz}. Additional scatter is added to $\dmc$ to account for sheets, filaments and voids in the IGM, as well as imperfect subtraction of Milky Way and host galaxy components.  We assume these components take the form of zero mean Gaussian noise, with $\sigma_\mathrm{IGM}=10$~\pccm~\cite{2018ApJ...852L..11S},  $\sigma_\mathrm{MW}=15$~\pccm~\cite{2015MNRAS.451.4277D}, and  $\sigma_\mathrm{host}=50$~\pccm~\cite{2018arXiv180401548W, 2019ApJ...872...88R}, which we add in quadrature  to give $\sigma=53$ \pccm.

Noting that the CGM component produces long high-DM tails in $P(\dmc | z)$, which are relatively well approximated by log-normal distributions, we decide to fit Gaussian distributions to $ \ln\dmc$, and constrain parameters using $\ln \dmc$ data, with Gaussian errors. However, to avoid the subtraction residuals generating negative values of $\dmc$ near $z=0$ (which would result in $\ln\dmc$ being undefined), we add a constant offset to $\dmc$. We thus define the quantity
\begin{align}
X(z) \equiv \ln(\dmc(z) + C), \label{defX}
\end{align}
and choose  $C = 200$ \pccm. We divide the simulated $X(z)$ data into $10^2$ redshift bins in the range $0\leq z \leq 3$, and in each bin, compute the histogram density and fit a Gaussian distribution (i.e. fit for mean and variance for each bin). We then associate the variance in each bin with the redshift of the bin centre, and fit a function of the form
\begin{align}
\sigma^2_{X}(z) =  a e^{ b z } + c e^{d z },  \label{varfit}
\end{align}
where $z$ is the redshift of the bin centre, and  $\{a,b,c,d\}$ is the set of fitting parameters.  Finally, we simulate mock catalogues of data by promoting $X$ to a random variable, sampling from a normal distribution given by
\begin{align}
X(z) \sim  \mathcal{N}\left( \overline{X}(z) , ~\sigma^2_{X}(z)  \right) ,
\end{align}
where $\overline{X}(z)$ is given by Eq.~\er{defX}, with $\dmc=\dmcm$. To investigate the impact of sample size, we populate catalogues with $\nfrb=\{10^2, 10^3\}$. We do not investigate the effect of redshift distribution, and in all cases assume the comoving number density of sources is constant, with a minimum luminosity cutoff. We thus model the distribution of sources as \cite{2018PhRvD..98j3518M}
\begin{align}
p(z) \propto \frac{\chi(z)^2}{(1+z)H(z)} e^{-d_L^2(z)/d_L^2(z_\mathrm{cut})},
\end{align}
where $\chi(z)$ is the radial comoving distance, $d_L(z)$ is the luminosity distance, and $z_\mathrm{cut}$ is the redshift of the luminosity cutoff. For both values of $\nfrb$, we generate 50 realisations of the data, run the MCMC fit, and compare the resulting constraints. 

The distribution of $\dmc$ and $X$ data can be seen in the left panel of Fig.~\ref{P_DMz}. We plot histogram densities in the various redshift bins (coloured blocks) derived from simulating $N=10^4$ sightlines out to $\zlim=3$.  The corresponding best-fit Gaussian to the $X$ data is shown in the bottom panel (solid coloured lines). 
The right panel of Fig.~\ref{P_DMz} shows the model for $\fd(z)$ (top), and it impact on the simulated data (below). A single realisation of the $X$ and corresponding $\dmc$ data, with $\nfrb=10^2$ (black points), together with their mean values (red lines). The best-fitting parameters of Eq.~\er{varfit}  are shown in Table \ref{bestfit}. We find these values provide a good fit out to $z=3$.

\begin{table}[htb]
\begin{tabular}{c c}
\multicolumn{1}{l}{Parameter~~} &  ~Best-fit  \\ \hline \hline
{\boldmath$ a$}                              & 0.0274     \\
{\boldmath $b$}                              & -0.5878    \\
{\boldmath $c$}                              & 0.0603     \\
{\boldmath $d$}                              & -3.4974
\end{tabular}
\caption{Best fitting parameter values which describe the variance of $X(z)$.}
\label{bestfit}
\end{table}

\section{Parameter Estimation}
In the MCMC analysis we fit  $X$ data using the $\chi^2$ statistic as a measure of likelihood for the parameter values, with log-likelihood function given by
 \begin{align}
 \ln \mathcal{L_{\mathrm{FRB}}}(\theta|{\rm D}) = -\frac{1}{2} \sum_i \frac{\left( X_{i} - \overline{X}(z_i)  \right)^2}{ \sigma_{X,i}^2  } , \label{likelihood}
 \end{align}
where $\theta$ is the set of fitting parameters, ${\rm D}$ is the FRB data, and the sum over $i$ represents the sequence of FRB data in the sample. We use the Python package {\tt emcee} \cite{2013PASP..125..306F} to generate the MCMC chains, and {\tt GetDist}\footnote{{\tt GetDist} is available at \url{https://github.com/cmbant/getdist}} for plotting and analysis.  

To assess how well FRBs might help to constrain $\fd(z)$, we consider two different parametric models. Firstly, a single fixed constant,  
\begin{align}
\fd(z)= \fdb, \label{fconst}
\end{align} 
where $\fdb$ represent a weighted average of the diffuse gas fraction in the redshift range or interest. And secondly, and a two parameter model given by
\begin{align}
\fd(z) = f_0 + f_a \frac{z}{1+z}, \label{f0fa}
\end{align}
where $f_0$ is the value of $\fd$ today, and $f_a$ is its derivative with respect to the scale factor, $a(t)$. 
We then fit for
\begin{align}
\vec{\theta} = \left( \omb h^2, \omm, H_0, \vec{f} ~\right), \label{params}
\end{align}
where $\vec{f}$ contains the parameters associated with the relevant $\fd(z)$ model, namely
\begin{align}
\vec{f} =  \begin{cases}
       \fdb  &\text{where Eq.~\er{fconst} applies,} \\
       (f_0, f_a)  &\text{where Eq.~\er{f0fa}  applies.} \\ 
     \end{cases}
\end{align}
To forecast the combined constraints, FRB+CBSH, we extract the relevant parameter covariance matrix from the chains provided by the {\it Planck} 2016 data release, and include it as a prior in the analysis.  The log-prior is given by
\begin{align}
\ln P(\theta)  = -\frac{1}{2} \xi \mathbf{C}^{-1} \xi,
\label{lnP}
\end{align}
where $P(\theta)$ is the prior probability associated with the set of parameter values $\theta$, $\mathbf{C}$ is the covariance matrix,  and $\xi = \theta-\tfid$ is the displacement in parameters space between the relevant parameter values and the fiducial values.

\section{Results}
\label{sec:result}

 \begin{figure*}[thb]
 \centering
 \begin{overpic}[width=.65\textwidth]{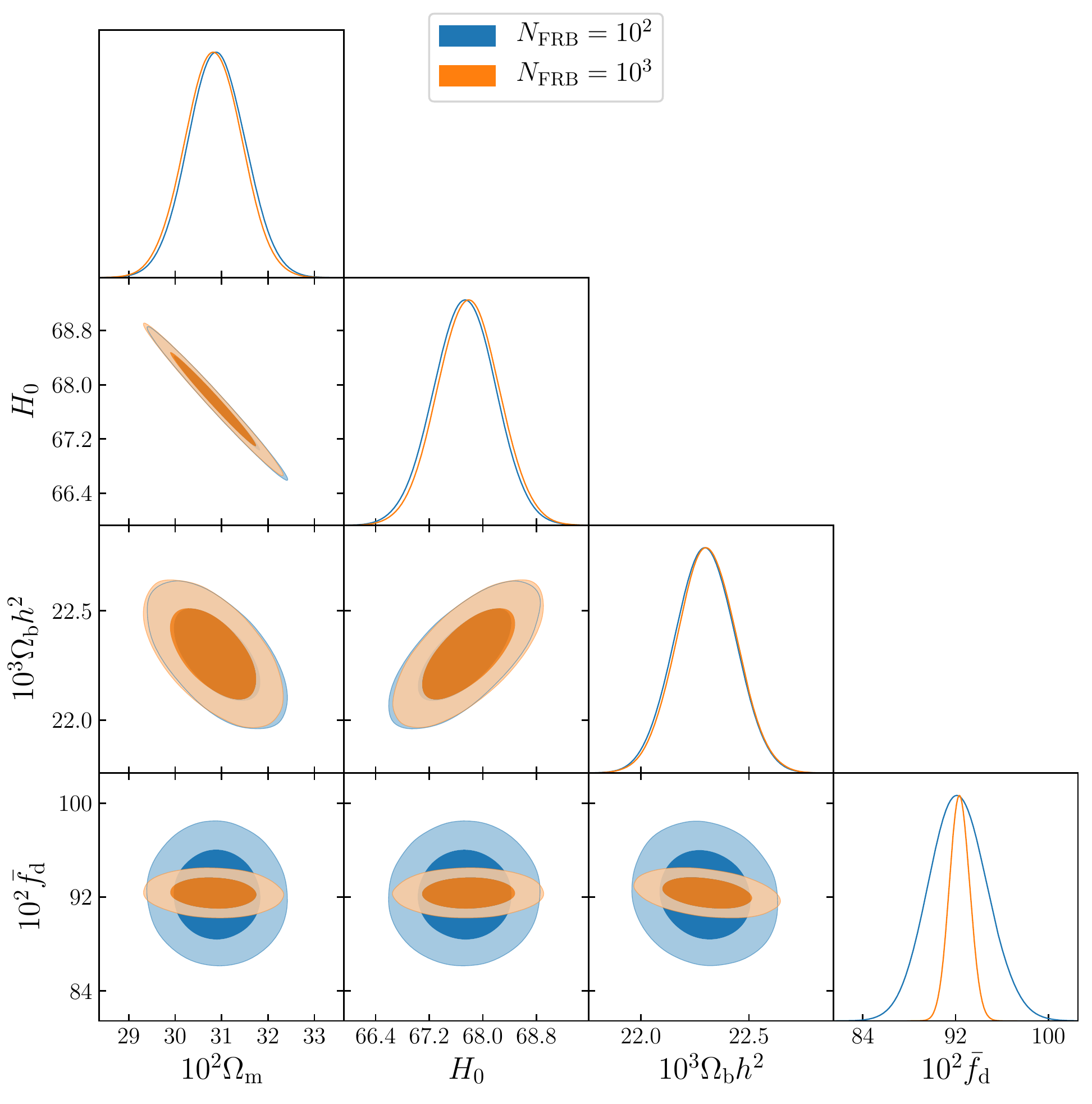}
 \put(60,53){\includegraphics[width=.4\textwidth]{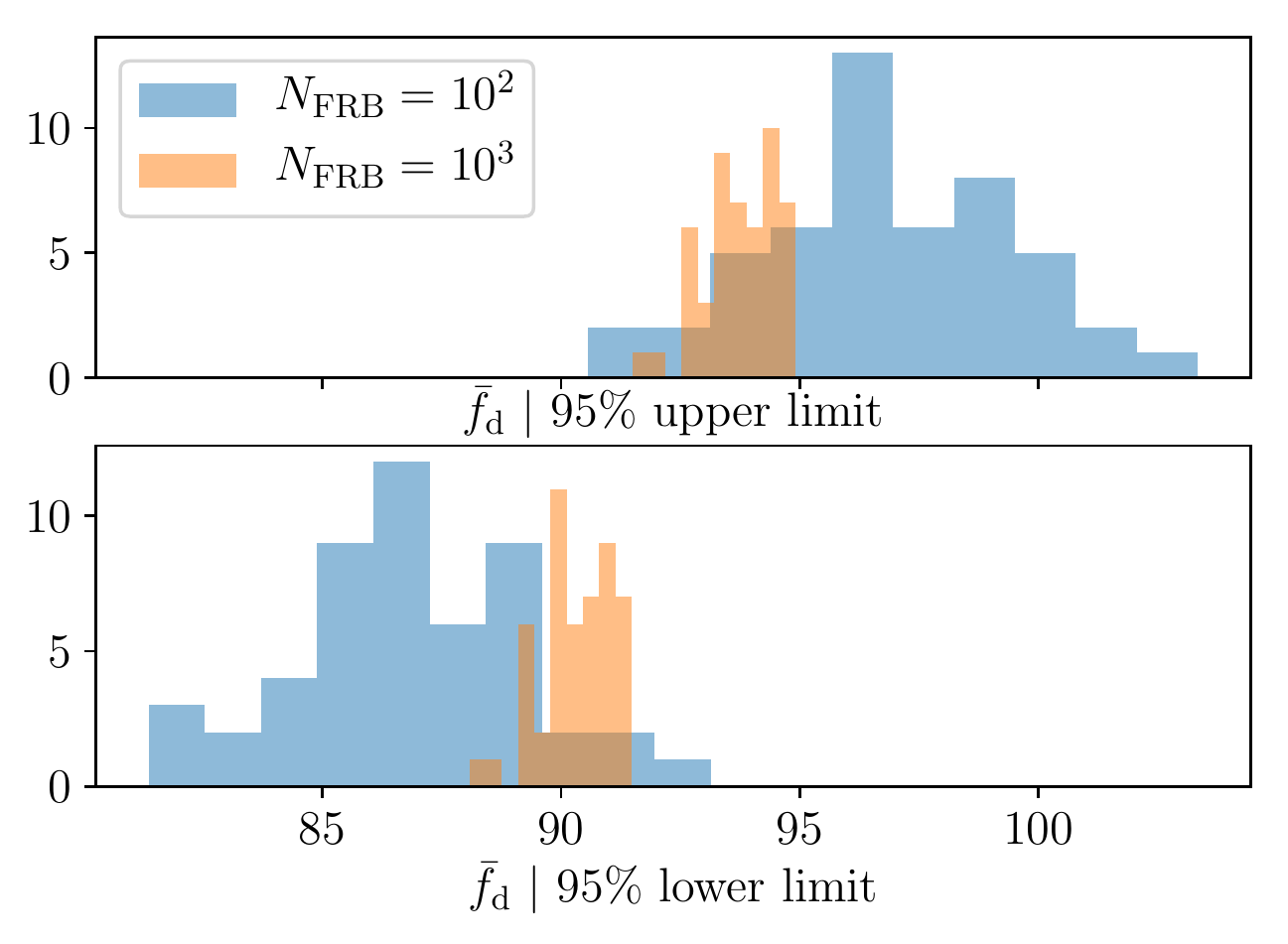}}
 \end{overpic}
\caption{Triangle plot of the marginalised posteriors parameter constraints, in the case where we it for $\fd(z)=\fdb$. Orange contours correspond to catalogues with $\nfrb=10^2$, and blue contours to catalogues with $\nfrb=10^3$. The grey shaded area corresponds to forbidden regions where $\fdb>0$. The CBSH priors on the cosmological parameters are coincident with the orange and blue contours, and so are not shown here. Inset: Histograms of the 95\% upper and lower bounds on $\fdb$ derived from 50 different realisations of the data.}
\label{triangle_base_f_varN}
\end{figure*}

\begin{table}
\begin{tabular} { l  c c}
$\nfrb$ & $10^2$ & $10^3$ \\
\hline\hline
 Parameter &  95\% limits~~~~~\\
\hline
{\boldmath$H_0            $} & $67.73^{+0.90}_{-0.91}     $& $67.79^{+0.90}_{-0.90}     $	\\
{\boldmath$10^3 \Omega_{\mathrm b} h^2$ }& $22.30^{+0.27}_{-0.27}      $&	$22.30^{+0.27}_{-0.27}     $\\
{\boldmath$10^2 \Omega_{\mathrm m}   $ }& $30.9^{+1.2}_{-1.2}        $&	$30.8^{+1.2}_{-1.2}       $\\
{\boldmath$10^2 \bar{f}_\mathrm{d}   $ }& $ 92.2^{+5.0}_{-4.9}                     $&	$92.3^{+1.7}_{-1.7}        $	\\
\hline 
\end{tabular}
\caption{Typical constraints on the base model parameters, for both values of $\nfrb$ considered here.}
\label{base_f_table}
\end{table}

We find that the posterior cosmological constraints show no improvement over the CBSH priors, even with $\nfrb=10^3$. However, the constraints on $\fdb$ appear promising. This can be seen in Fig.~\ref{triangle_base_f_varN}, where we show a triangle plot of all marginalised posterior constraints derived from a typical realisation of the data, together with histograms of the 95\% upper and lower bounds of $\fdb$ from all 50 realisations. Blue contours correspond to catalogues with $\nfrb=10^2$, and orange to catalogues with $\nfrb=10^3$. The corresponding $95\%$ limits on the fitting parameters are shown in Table~\ref{base_f_table}.

In general we find the catalogues with  $\nfrb=10^2$ all produce constraints on $\fdb$ of a few percent. A typical realisation of the data gives $10^2 \fdb=92.2^{+5.0}_{-4.9}$ at $95\%$ confidence. And the scatter in the posterior constraints on $\fdb$, across all realisations of the data is $\sim10\%$, as shown by the blue histogram at the top right of Fig.~\ref{triangle_base_f_varN}. It should be noted that this variation in the posterior constraint on $\fdb$ is not only due to intrinsic sample variance, but also as a result of fitting a single constant to and evolving function.  When the catalogue size increased to $\nfrb=10^3$ we find the data produce sub-percent constraints on $\fdb$, with a typical realisation of the data giving $10^2\fdb=92.3^{+1.7}_{-1.7}$ at $95\%$ confidence. Across all realisations we find the scatter in the posterior constraint is reduced to $\lesssim 5\%$. In all cases, we find no appreciable change in the posterior constraints when Milk Way and Host Galaxy subtraction residuals are set to zero. Doing so only tends to further skew the distribution $P(\dmc|z)$, making the Gaussian approximation to $P(X|z)$ less appropriate. 
 
 \begin{figure*}[thb]
 \centering
\includegraphics[width=.44\textwidth]{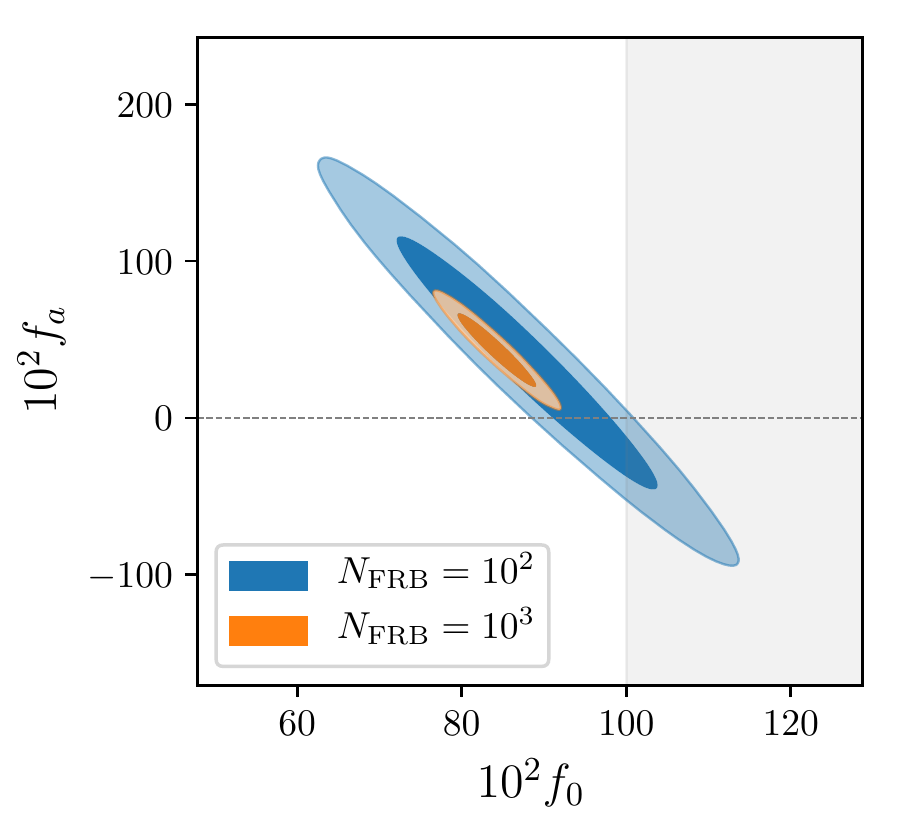}
\includegraphics[width=.51\textwidth]{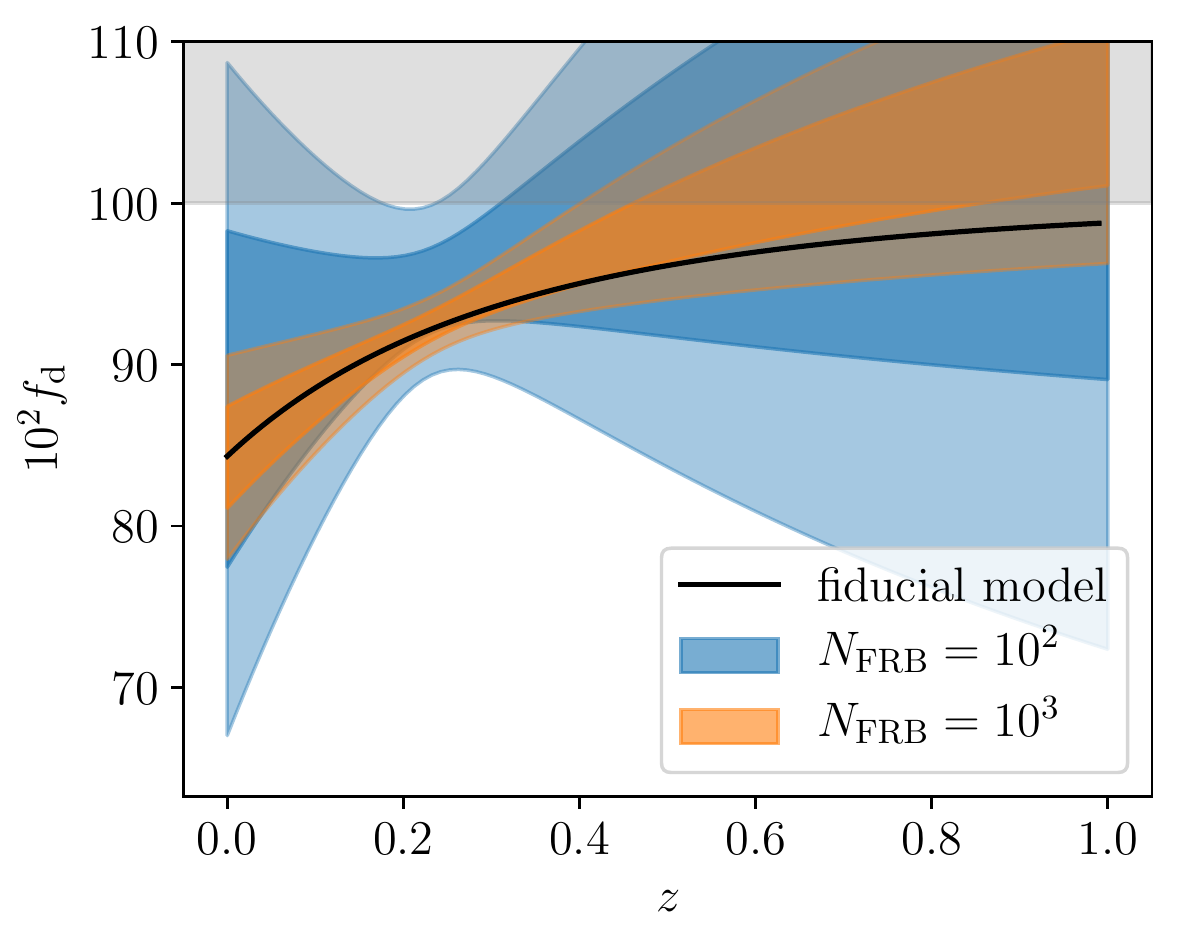}
\caption{Posterior constraints on the two parameter model for $\fd(z)$,  for both values of $\nfrb$ considered here. Left panel shows shows the 2d constrain contours for  $(f_0,f_a)$. The right panel shows the fiducial model  for $\fd(z)$ used when simulating the data (black line),  together with the reconstructed 1- and 2-$\sigma$ confidence intervals. Blue contours correspond to $\nfrb=10^2$, and orange contours to $\nfrb=10^2$. Grey shaded areas indicate the forbidden region where $\fd(z) > 1$.}
\label{base_f0fa}
\end{figure*}

Constraints on the two parameter model given by Eq.~\er{f0fa} are shown in Fig.~\ref{base_f0fa}. The left panel shows a 2d contour plot of the marginalised posterior constraints in the $(f_0,f_a)$ plane. The right panel shows the $1-$ and $2-\sigma$ confidence intervals of the reconstucted function $\fd(z)$, together with the fiducial model used in modeling the data. Blue contours correspond to catalogues with $\nfrb=10^2$, and orange to catalogues with $\nfrb=10^3$. The shaded grey areas indicate the forbidden regions where $\fd(z)>1$. The corresponding marginalised $95\%$ limits on $(f_0, f_a)$ are shown in Table~\ref{base_f_table}. The cosmological constraints and prior dominated, and so not shown here.

With $\nfrb=10^2$ we find no evidence for the evolution of $\fd(z)$. This can be seen in the left panel, where the blue contours are entirely consistent with $f_a=0$. However, increasing the size of the catalogue to $\nfrb=10^3$ allows one to detect the evolution in $\fd(z)$. We find a typical realisation of the data produced $10^2 f_a =43^{+30}_{-30} $ at 95\% confidence. This suggests that if $\fd(z)$ evolves in the redshift range of interest, as numerical simulations suggest it does, it should be possible to constrain its evolution with this method. 

\begin{table}
\begin{tabular} { l  c c}
$\nfrb$ & $10^2$ & $10^3$ \\
\hline\hline
 Parameter~~&  95\% limits\\
\hline

{\boldmath $10^2 f_0                  $ }& $ 88^{+20}_{-20}                    $ & $ 84.3^{+6.2}_{-6.1}                    $ \\

{\boldmath $10^2 f_a                  $ }& $ 36^{+100}_{-100}                    $ & $ 43^{+30}_{-30}                    $\\

\hline
\end{tabular}
\caption{Typical constraints on the two parameter model for $\fd(z)$ given by Eq.~(\ref{f0fa}), for both values of $\nfrb$ considered here.}
\label{fpw3_table}
\end{table}

\section{Conclusions}
\label{sec:conclude}
As a new generation of radio telescopes begin to operate and take data, and the catalogue of FRBs inevitably grows, we are presented with the possibly of extracting additional information about our cosmos. In this paper we have investigated how future measurements of the \dmz relation coming from FRBs, might help to improve current cosmological constraints, and possibly inform the missing baryon problem. To this end,  we simulate mock FRB \dmz observations, and constrain parameters using MCMC techniques.

We have paid particular attention to modeling the scatter in the \dmz data that is expected due to the CGM of intervening galactic halos,  fluctuations in the IGM (sheets, voids, filaments), as well as the effect of imperfect subtraction of the Milky Way and FRB host galaxy DM. Combining all these sources of uncertainty together, we have found that the distribution of DMs is reasonably well approximated by a log-normal distribution. We thus simulated mock $X\equiv\ln( DM + C)$ data with Gaussian distributed errors, and use that to estimate parameters. In addition, we have provided fitting formulae for the total variance in the $X$ data, valid out to $z=3$, and find no appreciable change to the best-fit when Milky Way and host galaxy residuals are  zero. The scatter due to intervening galactic halos dominates. There is the possibility, however,  that a more skewed distribution will provide a better fit to the data when Milky Way and host galaxy residuals are set to zero, which could be used to construct a better-fitting likelihood function.  We leave this to be investigated in future work.

While previous work has shown that MCMC fitting of  \dmz data may help to improve on the current CBSH constraints, we find this is not the case when $\fd(z)$ is allowed to evolve (as observations and simulations suggest it does), and any additional unconstrained parameters which describe the diffuse gas fraction are included in the fit. Indeed, current observational constraints on $\fd(z)$ are quite poor, many times weaker than the cosmological constraints, and so it is not surprising that cosmological constraints do not improve when an additional unconstrained parameter is introduced. We do however find promising constraints on  $\fd(z)$. 

From a catalogue of just $10^2$ \dmz data, in the range $0 \leq z \lesssim 1$, we find a typical constraint on the mean diffuse gas fraction, $\fdb$, of a few percent. And with $\nfrb=10^3$, a sub-percent constraint on $\fdb$. Indeed, such a detection would alleviate the missing baryon problem. This highlights the importance of sufficiently localising FRBs so that they can be associated with a host galaxy, and the importance of developing methods to remove the host galaxy contribution to the observed DM. Furthermore,  it may be possible to detect the redshift evolution of $\fd(z)$ using a two parameter model. This would allow for comparison to be made with the predictions of numerical simulations, and may help to constrain models of large-scale structure~\cite{McQuinn14}. We leave this as a  possibility to investigate in future work.

\begin{acknowledgments}
We thank J. Xavier Prochaska for help using the FRB packages, as well as Kavilan Moodley and Kurt van der Heyden for useful discussion. A  Walters acknowledges support from  the National Research Foundation of South Africa (NRF) [grant  numbers 105925,  110984, 109577]. Y-Z.M. acknowledges the support by National Research Foundation of South Africa (No. 105925, 109577), and UKZN Hippos cluster. A. Weltman gratefully acknowledges support from the South African Research Chairs Initiative of the Department of Science and Technology and the NRF.  
\end{acknowledgments}

\bibliographystyle{IEEEtran}
\bibliography{frb_refs}

\end{document}